\documentclass[12pt]{article}

\usepackage{amsfonts}
\usepackage{amssymb}
\usepackage{amsmath}
\usepackage{amsthm}
\usepackage{eurosym}
\usepackage{upgreek}
\usepackage{tikz}
\usepackage{graphicx}
\usepackage{multirow}
\usepackage{subfigure}
\usepackage{pgfplots}
\usepackage{pgfmath}
\usepackage{stmaryrd}
\usepackage{units}
\RequirePackage[colorlinks,citecolor=blue,linkcolor=red,urlcolor=blue]{hyperref}
\RequirePackage{natbib}

\usetikzlibrary{shapes,snakes}

\theoremstyle{theorem}
\newtheorem{theorem}{Theorem}
\newtheorem{proposition}{Proposition}

\newtheorem{exam}{Example}

\newtheorem*{exam1}{Example 1 (cont)}

\newtheorem*{exam2}{Example 2 (cont)}

\newtheorem*{exam4}{Example 4 (cont)}

\newtheorem{assum}{Assumption}

\newtheorem{rema}{Remark}
\newenvironment{remark}{\begin{rema} \rm }{\hfill $\triangleleft$ \end{rema}}
\newtheorem{defin}{Definition}
\newenvironment{definition}{\begin{defin} \rm }{\hfill $\triangleleft$ \end{defin}}

\DeclareFontFamily{U}{mathx}{\hyphenchar\font45}
\DeclareFontShape{U}{mathx}{m}{n}{
      <5> <6> <7> <8> <9> <10>
      <10.95> <12> <14.4> <17.28> <20.74> <24.88>
      mathx10
      }{}
\DeclareSymbolFont{mathx}{U}{mathx}{m}{n}
\DeclareMathSymbol{\bigtimes}{1}{mathx}{"91}

\makeatletter

\renewcommand*{\@seccntformat}[1]{%
  \csname the#1\endcsname.\quad}
\makeatother

\begin{document}

\title{A robust measure of complexity}

\author{
\textsc{Egor Bronnikov}\footnote{Homepage: \url{egorbronnikov.github.io}; E-mail: \href{mailto:egor.bronnikov@maastrichtuniversity.nl}{\texttt{egor.bronnikov@maastrichtuniversity.nl}}}  \ \ \ \ \ \ \textsc{Elias Tsakas}\footnote{Homepage: \url{www.elias-tsakas.com}; E-mail: \href{mailto:e.tsakas@maastrichtuniversity.nl}{\texttt{e.tsakas@maastrichtuniversity.nl}}} 
\\
\small{\textit{Maastricht University}} 
}

\date{January, 2025 \\ \vspace{2mm} \small{Preliminary draft} \\ 
\vspace{2mm} \href{http://www.elias-tsakas.com/Research/Papers/RobustComplexity.pdf}{Latest version of the paper.}}

\maketitle

\begin{abstract}

\noindent We introduce a robust belief-based measure of complexity. The idea is that task A is deemed more complex than task B if the probability of solving A correctly is smaller than the probability of solving B correctly regardless of the reward. We fully characterize the corresponding order over the set of tasks. The main characteristic of this relation is that it depends, not only on difficulty (like most complexity definitions in the literature) but also on ex ante uncertainty. Finally, we show that for every task for which information is optimally acquired, there exists a more complex task which always induces less effort regardless of the reward.

\vspace{0.5\baselineskip}



\end{abstract}

\section{Introduction}

Complexity is a fundamental concept across numerous scientific domains, including computer science, cognitive sciences, neuroscience, etc. More recently, its importance has also been recognized by (behavioral) economists, as it is a key determinant of decisions in many settings, and more importantly it has the potential to explain mistakes that people systematically make in such decisions  \citep{BanovetzOprea2023, EnkeGraeberOprea2024, Oprea2024}. 

At the same time, although most notions of complexity somehow reflect the difficulty to handle a task ---or the associated cost--- it is also the case that complexity is a term which is traditionally used in a casual way, without consensus on what exactly it means \citep{GabaixGraeber2024}. This lack of a widely accepted precise definition can possibly explain why scholars have also focused on measuring complexity, which can be in turn serve as a proxy for different types of complexity \citep{OpreaReview2024}.\footnote{Measurements and formal definitions of complexity are closely linked with each other, in an analogous way belief elicitation mechanisms \citep{Brier1950, Savage1971} are linked with formal notions of subjective probability became established \citep{Savage1954, AnscombeAumann1963}.} Common ways to measure complexity include direct metrics \citep{Oprea2020}, behavioral metrics \citep{BanovetzOprea2023}, and most importantly for this paper belief-based metrics \citep{EnkeGraeber2023, EnkeGraeberOprea2024, EnkeGraeberOpreaYang2024}. 

The key idea within this last stream of literature is to use an agent's belief about their own accuracy as a proxy for complexity. This is also consistent with theoretical results which show that people are more accurate when solving simpler tasks \citep{Goncalves2024}. However, as appealing as this approach is, there is a serious caveat, which stems from the fact expected accuracy depends both on complexity as well as on the effort to handle the task. And since effort depends non-linearly on the underlying reward (for correctly solving the task), it often happens that expected accuracy is sensitive to the reward. And this naturally gives rise to the question: if different rewards yield different revealed complexity order over the set of tasks, which one should we use?

In this paper, we take a robust approach, proposing the following way to rank tasks: we say that task A is revealed to be more complex than task B, whenever the chances to correctly solve A are smaller than the chances to correctly solve B, \textit{for every (extrinsic) reward}. Obviously, this is a rather conservative criterion, as it imposes a rather strong dominance condition. But at the same time, whenever satisfied, it provides quite strong and convincing evidence that the tasks are indeed ranked in this way, thus indirectly providing a sufficient condition that every reasonable definition of complexity should satisfy.

The main result of the paper provides a full characterization of the complexity order that the aforementioned criterion induces (Theorem \ref{T:main theorem incompleteness}). Not surprisingly, the complexity order is incomplete. But they key insight is that it depends on two distinct parameters, viz., the difficulty of the task and the ex ante uncertainty.  More specifically, higher difficulty is only a necessary condition for higher complexity. In order to also become sufficient, the agent cannot be ex ante much more uncertain about the state realization. For example, in order for a student to find exam A more complex than exam B, it must be both the case that A is more difficult, and moreover that the student is not much better prepared about B. This is a novel channel of complexity, as in most of the existing literature, complexity is typically associated only with some sort of difficulty. 

The fact that complexity has these two distinct dimensions allows us to establish a link with the decision-theoretic literature on incomplete preferences. In particular, we show that our complexity order is represented by a vector (utility) function, like in \cite{Ok2002}. 

This representation allows us to do comparative statics with respect to (unobservable) intrinsic incentives that the agent may have on top of the extrinsic reward. In particular, we show that increasing the intrinsic incentives has a dual effect, viz., on the one hand, it leads to a more complete order (i.e., we can compare more tasks with one another), but at the same time, it also makes the differences between tasks less salient (i.e., it is less clearcut which task is more complex). In this sense, 

Finally, we go back to the relationship between complexity and induced effort, where effort is seen as a proxy for the cost that the agent optimally incurs when handling the task. As it has been already pointed out in the literature, we should not expect effort to be monotonic with respect to complexity, i.e., it might be the case that simpler tasks induce more effort, as the agent may not find it worthwhile to try at all in order to answer a task which is anyway too difficult \citep{Goncalves2024}. However, this argument is typically obtained for fixed rewards. Thus, we naturally ask whether it is robust across different rewards. 

Our second main result shows that this is indeed the case, i.e., for every task against which the agent optimally acquires some information, there exists some more complex task which always induces less effort, regardless of the size of the reward (Theorem \ref{T:monotonicity I}). The result is quite surprising, as one would expect that for sufficiently high reward, the agent would put maximum effort both for the complex and the simple task. However, what happens here is that the more complex task that we find is a but more difficult, and crucially involves much more ex ante uncertainty. As a result, the role of increased difficulty is relatively small.


The literature on complexity is vast, and as such we are de facto forced to make a selection of what in our view is the most relevant subset. 

Early work focused primarily on the role of strategy complexity within game theory \citep{Rubinstein1986, AbreuRubinstein1988}. More recently,  the focus has shifted towards explaining mistakes and irrationalities, e.g.,  \cite{Oprea2024} study the effect of complexity on risk preferences, and \cite{EnkeGraeberOprea2024} the respective effect on time preferences. There is also interest in formalizing definitions of complexity, e.g.,  \cite{GabaixGraeber2024} build a general model of production within a cognitive economy in order to operationalize complexity, whereas \cite{OpreaReview2024} borrows insights from computer science to introduce a framework within which complexity reflects the cost for handling a task. Others, define it as the signal-to-noise ratio \citep{Goncalves2024}, similarly what is often done in psychometrics. The common denominator throughout most of this literature is that definitions of complexity are typically input-based, i.e., they somehow reflect the underlying difficulty to process and handle a task. 

What is closer to our work is the literature on measuring complexity. As  \cite{OpreaReview2024} elegantly points out, this literature can be classified into three large streams, depending the measurement tool. Within the first stream, we encounter measurement with direct metrics, such as willingness to pay in order to avoid dealing with a certain task \citep{Oprea2020}, response times \citep{Goncalves2024}, and biometrics \citep{vanderWelvanSteenbergen2018}. The second stream leverages behavioral metrics, such as procedural measurements \citep{BanovetzOprea2023}, and choice inconsistencies \citep{Woodford2020}. 

Finally, the third stream, within which our paper belongs, uses belief-based metrics. These include subjective rankings, like for instance in \cite{GabaixGraeber2024} where subjects are simply asked to rank tasks with respect to complexity, and beliefs about optimality of the subject's own accuracy \citep{EnkeGraeber2023, EnkeGraeberOprea2024, EnkeGraeberOpreaYang2024}, like in our paper. Of course, in all  aforementioned papers, rewards are fixed. The effects of varying rewards are discussed in \citep{AlaouiPenta2022}.

This entire literature is part of a surging field of Cognitive Economics \citep{Caplin2025, Enke2024}, which also incorporates topics such as rational inattention, cognitive uncertainty, etc.

The paper is structured as follows: In Section \ref{S:complexity} we introduce our measure of complexity and prove our main characterization result. In Section \ref{S:properties} we study properties of the complexity order which is induced by our measure. In Section \ref{S:complexity and effort} we study the relationship between our complexity measure and the induced effort.

\section{A measure of subjective complexity}\label{S:complexity}

Take a binary state space $\Theta=\{\theta_0,\theta_1\}$. An agent is asked to guess the realization of $\Theta$. Let $Y=\{y_0,y_1\}$ denote the possible results of this guessing task, i.e., $y_0$ denotes a wrong guess and $y_1$ denotes a correct guess. Let $X:=[x_0,\infty)$ be a convex set of monetary rewards for guessing correctly. In this sense, $Y$ is the set of intrinsic rewards, and $X$ is the set of extrinsic rewards. 

The agent's preferences over the set of acts $(X\times Y)^\Theta$ admit a SEU representation with Bernoulli utility function $u:X\times Y \rightarrow\mathbb{R}$, which is often abbreviated by $u_0(x):=u(x,y_0)$ and $u_1(x):=u(x,y_1)$. For both $y\in Y$, the utility function is continuously increasing and unbounded in $X$. We use the difference 
\begin{equation}
w:=u_1(x_0)-u_0(x_0)
\end{equation}
as a measure of intrinsic motivation, i.e., this is how much the agent cares about being right, in the absence of extrinsic rewards. We assume that $w\geq0$. Note that the previous inequality is only weak, i.e., in the absence of extrinsic incentives, the agent may or may not care about intrinsic incentives. Furthermore, observe that we do not impose any condition on the difference $u_1(x)-u_0(x)$ for any $x>x_0$. Neither do we assume separability of $u$ with respect to $X$ and $Y$. Finally, without loss of generality, the utility function is normalized so that $u(x_0,y_0)=0$. 

Going back to the guessing task, for some fixed extrinsic reward $x\in X$, the guess $r\in\Theta$ is an act which yields utility $u_1(x)$ at state $\theta=r$ and utility $0$ at $\theta\neq r$. Thus, the agent's indirect expected utility, as a function of the probability $q\in[0,1]$ that she attaches to $\theta_1$, is given by
 \begin{equation}\label{EQ:conditional accuracy}
g(q):=u_1(x)\max\{q,1-q\},
\end{equation}
which is obviously proportional to the probability she attaches to her best guess being correct.

Suppose that the agent holds a prior belief which assigns probability $p\in[0,1]$ to $\theta_1$. This belief incorporates her prior knowledge/experience, and therefore represents the degree of her ex ante uncertainty in terms of proximity to the (maximally uncertain) uniform belief:
\begin{equation}
\phi := 1-2\big|p-\nicefrac{1}{2}\big|.
\end{equation}
That is, the larger $\phi$, the more ex ante uncertain the agent is, with $\phi=0$ meaning that the prior attaches probability 1 to one of the two states, and $\phi=1$ meaning that $p$ is uniformly distributed. This notion of uncertainty resembles the usual measures of uncertainty from information theory \citep{CoverThomas2006}. Then, before making a guess, the agent decides how much attention to pay. Attention is modelled with a Bayesian signal, which is uniquely identified by a mean-preserving distribution of posterior probabilities \citep{KamenicaGentzkow2011}. The set of all signals is denoted by 
$$\Pi(p)= \ \Bigm\{\pi\in\Delta\bigl([0,1]\bigr):\mathbb{E}_\pi(q)=p\Bigm\}.$$
For any signal $\pi\in\Pi(p)$, define the agent's ex ante indirect expected utility,
\begin{equation}
G(\pi):=\mathbb{E}_\pi\bigl(g(q)\bigr).
\end{equation}
It is not difficult to see that $G(\pi)$ is proportional to the probability $G(\pi)/u_1(x)$ that she attaches ex ante (i.e., before $\pi$ is realized) to her best guess being eventually correct.

Aligned with the rational inattention literature, we assume that attention is costly. The cost function is assumed to be uniformly posterior separable \citep{Caplin2022}, i.e., there is a strictly convex function $c:[0,1]\rightarrow\mathbb{R}$ such that the cost of signal $\pi\in\Pi(p)$ is given by
\begin{equation}
C(\pi)=\kappa\Bigl(\mathbb{E}_{\pi}\bigl(c(q)\bigr)-c(p)\Bigr),
\end{equation}
where $c(q)$ represents the agent's marginal cost for acquiring information, and $\kappa>0$ is a parameter of the task's difficulty. Note that consistently with the complexity literature \citep{OpreaReview2024}, the cost consists an objective part (viz., the parameter $\kappa$) and a subjective part (viz., the function $c$). Such cost functions have solid foundations \citep{Denti2022} and are supported by experimental evidence \citep{DeanNeligh2023}. Throughout the paper, we will focus on symmetric cost functions, which include the Shannon entropy \citep{Sims2003}, the Shorrocks entropy \citep{Shorrocks1980}, the Tsallis entropy \citep{Caplin2022} as special cases. Formally, this means that for every $q\in[0,1]$, we have $c(q)=c(1-q)$. For an axiomatization of symmetric cost functions, see \cite{HebertWoodford2021}. 

The agent faces a tradeoff, in that more informative signals help her to achieve higher expected utility, but at the same time are also more costly. That is, formally speaking, the agent solves the following optimization problem:
\begin{equation}\label{EQ:general optimization problem}
\max_{\pi\in\Pi(p)}\Bigl(G(\pi)-C(\pi)\Bigr).
\end{equation}
In the previous optimization problem, four parameters enter the picture: the extrinsic reward $x$, the intrinsic incentives $w$, the ex ante uncertainty $\phi$, and the difficulty $\kappa$. The only one which is usually observable by the analyst and can potentially vary is the extrinsic reward $x$. The intrinsic reward $w$ is typically unobservable, but most importantly cannot vary, as it simply reflects the agent's exogenous satisfaction for solving a task correctly. In this sense, throughout the paper, it is treated as an individual characteristic of the agent. Finally, the ex ante uncertainty $\phi$ and the difficulty $\kappa$ are inherently linked with the task itself, and in particular they represent the past information and the cost for acquiring new information, respectively. In this sense, we will henceforth treat the pair $(\phi,\kappa)$ as a description of the task, and correspondingly  
$$\mathcal{T}:=[0,1]\times (0,\infty)$$ 
as the set of all tasks. 

It is not difficult to verify that for every tuple of parameters $(x,w,\phi,\kappa)$ there is a unique solution to the optimization problem (\ref{EQ:general optimization problem}), henceforth denoted by $\pi(\cdot|x,w,\phi,\kappa)$. It follows from standard arguments \citep[e.g.,][]{KamenicaGentzkow2011,Matejka2015} that the optimal signal is given by concavifying the function $g(q)-\kappa c(q)$, as illustrated in the figure below. 

\begin{figure}[h!]
\begin{center}
\begin{tikzpicture}[scale=6]
\draw[line width=0.8,->] (0,0) -- (1.1,0);
\draw[line width=0.8,->] (0,-0.5) -- (0,0.8);
\draw[line width=0.8, dotted] (1,-0.5) -- (1,0.8);
\draw[red,line width=0.8,domain=0:1] plot(\x,{-\x+\x^2}) node[below right] {\footnotesize{$\kappa c(q)$}};
\draw[line width=0.8,dotted,green] (0,0) -- (0.5,0.25) -- (1,0);
\draw[line width=0.8,green] (0,0.5) -- (0.5,0.25) -- (1,0.5) node[right] {\footnotesize{$g(q)$}};
\draw[line width=0.8,green,dashed] (0.5,0.25) -- (0.5,0) node[below] {\footnotesize{$\nicefrac{1}{2}$}};
\draw[line width=0.8,domain=0.25:0.5] plot(\x,{0.5+0.5*\x-\x^2});
\draw[line width=0.8,domain=0.5:0.75] plot(\x,{1.5*\x-\x^2});
\draw[line width=0.8,dashed] (0.25,9/16) -- (0.25,0) node[below] {\footnotesize{$\delta$}};
\draw[line width=0.8,dashed] (0.75,9/16) -- (0.75,0) node[below] {\footnotesize{$1-\delta$}};
\draw[blue,line width=1,domain=0:0.25] plot(\x,{0.5+0.5*\x-\x^2});
\draw[blue,line width=1,domain=0.75:1] plot(\x,{1.5*\x-\x^2});
\draw[blue,line width=0.8] (0.25,9/16) -- (0.75,9/16);
\end{tikzpicture}
\end{center}
\vspace{-1\baselineskip}
\caption{\footnotesize{The green solid piecewise linear function is the net expected utility from guessing correctly, and the red line is the marginal cost for acquiring information. The blue curve is the concave closure of the difference $g(q)-\kappa c(q)$. The optimal signal distributes all the probability between the posteriors $\delta$ and $1-\delta$ whenever ex ante uncertainty is large (i.e., whenever $p\in(\delta,1-\delta)$), and it is completely uninformative otherwise.}}
\label{FIG:concavification}
\end{figure}
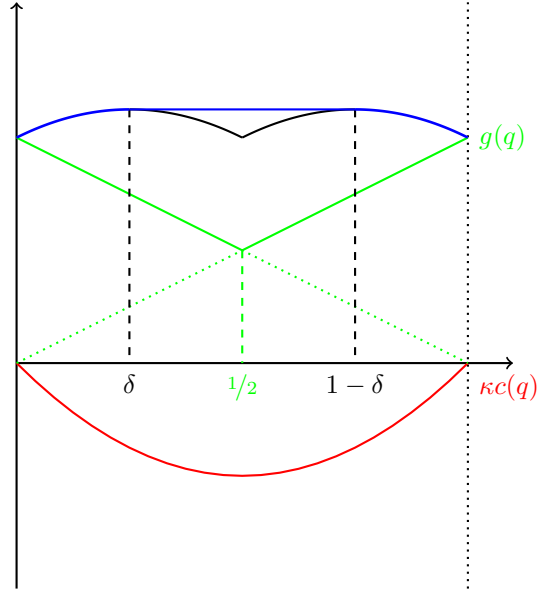

Then, the agent's subjective expected accuracy is given by
\begin{equation}
F(x,w,\phi,\kappa):=\frac{G\bigl(\pi(\cdot|x,w,\phi,\kappa)\bigr)}{u_1(x)}.
\end{equation}
This is equal to the expected probability of guessing correctly given that the optimal signal is used $\pi(\cdot|x,w,\phi,\kappa)$.

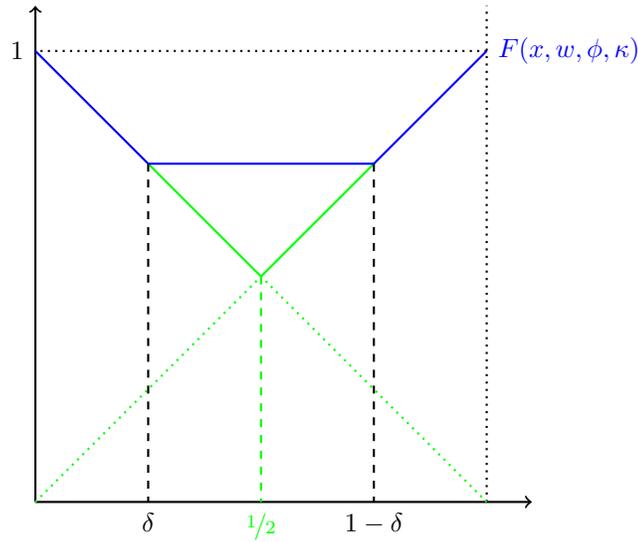
\begin{figure}[h!]
\begin{center}
\begin{tikzpicture}[scale=6]
\draw[line width=0.8,->] (0,0) -- (1.1,0);
\draw[line width=0.8,->] (0,0) -- (0,1.1);
\draw[line width=0.8, dotted] (1,0) -- (1,1.1);
\draw[line width=0.8, dotted] (1,1) -- (0,1) node[left] {\footnotesize{1}};
\draw[line width=0.8,dotted,green] (0,0) -- (0.5,0.5) -- (1,0);
\draw[line width=0.8,green] (0.25,0.75) -- (0.5,0.5) -- (0.75,0.75);
\draw[line width=0.8,green,dashed] (0.5,0.5) -- (0.5,0) node[below] {\footnotesize{$\nicefrac{1}{2}$}};
\draw[line width=0.8,dashed] (0.25,0.75) -- (0.25,0) node[below] {\footnotesize{$\delta$}};
\draw[line width=0.8,dashed] (0.75,0.75) -- (0.75,0) node[below] {\footnotesize{$1-\delta$}};
\draw[line width=0.8,blue] (0,1) -- (0.25,0.75) -- (0.75,0.75) -- (1,1) node[right] {\footnotesize{$F(x,w,\phi,\kappa)$}};
\end{tikzpicture}
\end{center}
\vspace{-1\baselineskip}
\caption{\footnotesize{The blue piecewise linear function is the expected accuracy, assuming that the optimal signal has been used.}}
\label{FIG:accuracy}
\end{figure}

Now, consistently with the existing literature which suggests that complexity is positively correlated with accuracy \citep{Goncalves2024}, we would like to use $F(x,w,\phi,\kappa)$ as a measure of subjective complexity. However, there is still a caveat: In our view, a sensible definition of complexity should be disentangled from the material reward that the agent will receive in case she solves the task. The analogy is simple. Subjective complexity should only be linked with characetristics of the task itself. In this sense, we will use subjective expected accuracy as a measure of subjective complexity only when it is robust with respect to the extrinsic reward, i.e., formally, we introduce the following definition.

\begin{definition}
For an agent whose intrinsic incentives are given by $w\geq0$, we say that task $(\phi,\kappa)$ is \textbf{\textit{subjectively more complex}} than task $(\phi',\kappa')$, and we write $(\phi,\kappa)\succeq_w(\phi',\kappa')$, whenever
\begin{equation}
F(x,w,\phi,\kappa)\leq F(x,w,\phi',\kappa')
\end{equation}  
for all $x\in X$. 
\end{definition}

The asymmetric and the symmetric parts of $\succeq_w$ are defined as usual. That is, we have $(\phi,\kappa)\succ_w(\phi',\kappa')$ whenever $(\phi,\kappa)\succeq_w(\phi',\kappa')$ and $(\phi,\kappa)\npreceq_w(\phi',\kappa')$, and respectively $(\phi,\kappa)\sim_w(\phi',\kappa')$ whenever $(\phi,\kappa)\succeq_w(\phi',\kappa')$ and $(\phi,\kappa)\preceq_w(\phi',\kappa')$. 

A task $(\phi,\kappa)$ is said to be trivial given intrinsic incentives $w$, if the optimal signal reveals the true state with certainty for every $x\geq x_0$, i.e., the support of the optimal signal $\pi(\cdot|x,w,\phi,\kappa)$ contains only posteriors that put probability 1 to a single state in $\Theta$. Obviously, if a task $(\phi,\kappa)$ is trivial, then it is simpler than any other task, i.e., $(\phi',\kappa')\succeq_w(\phi,\kappa)$ for all $(\phi',\kappa')\in\mathcal{T}$. The set of non-trivial tasks is henceforth denoted by $\mathcal{T}_w\subseteq\mathcal{T}$, and it is characterized by a difficulty threshold $\kappa_w\geq0$.

\begin{proposition}\label{P:trivial task}
For each $w\geq0$ there is some $\kappa_w\geq0$, such that the following are equivalent:
\begin{itemize}
\item[(i)]  $\kappa> \kappa_w$.
\item[(ii)] $(\phi,\kappa)\in\mathcal{T}_w$ for every $\phi\in[0,1]$.
\end{itemize}
Moreover, $\kappa_w$ is increasing in $w$.
\end{proposition}

The idea is quite simple: in order for a task to be non-trivial, the information costs must be sufficiently large to guarantee that the intrinsic incentives alone are not strong enough to always lead to a perfectly informative signal. In our main result below, we characterize how non-trivial tasks are ranked in terms of subjective complexity.

\begin{theorem}\label{T:main theorem incompleteness}
For every fixed $w\geq0$, there is an increasing function $\phi_w:(\kappa_w,\infty)\rightarrow[0,1]$ such that the following are equivalent for any pair $(\phi,\kappa)$ and $(\phi',\kappa')$ of non-trivial tasks: 
\begin{itemize}
\item[(i)] $(\phi',\kappa')\succeq_w(\phi,\kappa)$.
\item[(ii)] $\kappa'\geq\kappa$ and $\phi'\geq\min\{\phi_w(\kappa),\phi\}$.
\end{itemize}
\end{theorem}

Let us first explain what $\phi_w(\kappa)$ stands for. Suppose that there is no extrinsic reward, i.e., let $x=x_0$. Then, for a given level of intrinsic incentives $w\geq0$, the agent will acquire information if and only if her ex ante uncertainty satisfies the following inequality:
\begin{equation}\label{EQ:information constraint}
\phi>\phi_w(\kappa),
\end{equation}
i.e., the prior needs to lie sufficiently close to the uniform belief in order for the agent to optimally acquire information. Obviously, $\phi_w$ is continuously decreasing with $\phi_w(\kappa)\rightarrow0$ as $\kappa\rightarrow\kappa_w$, and $\phi_w(\kappa)\rightarrow1$ as $\kappa$ grows arbitrarily large. This means without sufficiently large costs, the agent will always acquire the perfectly informative signal regardless of her prior, whereas when the cost becomes infinitely large she will not acquire information regardless of her prior belief. Moreover, $\phi_w$ is increasing in $w$, i.e., the stronger the intrinsic incentives are, the easier it becomes to acquire information. 

Graphically, the previous result is illustrated in the Figures below: which correspond to two cases, i.e., the one where the information-acquisition constraint of (\ref{EQ:information constraint}) holds, and the one where it does not. In the figures, we have taken $c(q)=q^2-q$, which is subdifferentiable at the boundaries  of the unit interval, and therefore for small cost parameters the agent will optimally acquire the perfectly informative signal regardless of the size extrinsic reward. This explains why $\phi_w$ is initially constant at 0, and as a result there is a grey region which contains the trivial tasks. Such horizontal part would not have appeared if the cost function was for instance entropic, in which case a perfectly informative signal is never optimal.

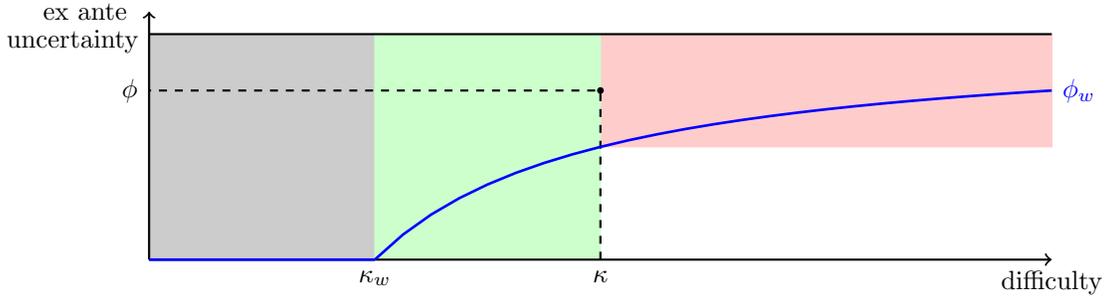
\begin{figure}[h!]
\begin{center}
\begin{tikzpicture}[scale=3]
\filldraw[opaque,white!80!red] (2,0.5) rectangle (4,1);
\filldraw[opaque,white!80!black] (0,0) rectangle (1,1);
\filldraw[opaque,white!80!green] (1,0) rectangle (2,1);
\draw[line width=0.8,->] (0,0) -- (0,1.1);
\draw (-0.05,1.1) node[left] {\footnotesize{ex ante}};
\draw (0,0.97) node[left] {\footnotesize{uncertainty}};
\draw[line width=0.8,->] (0,0) -- (4,0) node[below] {\footnotesize{difficulty}};
\draw[line width=0.8] (0,1) -- (4,1);
\draw[line width=0.8,dashed] (2,0) node[below] {\footnotesize{$\kappa$}} -- (2,0.75) -- (0,0.75) node[left] {\footnotesize{$\phi$}};
\draw (1,0) node[below] {\footnotesize{$\kappa_w$}};
\filldraw (2,0.75) circle(0.35pt);
\draw[blue,line width=1,domain=0:1] plot(\x,{0});
\draw[blue,line width=1,domain=1:4] plot(\x,{1-1/\x});
\draw[blue] (4,0.75) node[right] {\footnotesize{$\phi_w$}};
\end{tikzpicture}
\end{center}
\vspace{-1\baselineskip}
\caption{\footnotesize{This is the case when the ex ante uncertainty is large and therefore it is optimal to acquire information before making a guess. The red area contains the tasks that are deemed more complex than $(\phi,\kappa)$, and the green area are the tasks that are deemed simpler than $(\phi,\kappa)$. Finally, the grey area contains the trivial tasks, meaning that they are deemed simpler than every other task, including $(\phi,\kappa)$.}}
\label{FIG:main theorem binding information condition}
\end{figure}

Starting with the first case (Figure \ref{FIG:main theorem binding information condition}), we take a task $(\phi,\kappa)$ such that $\phi>\phi_w(\kappa)$. Then, the red region contains the tasks $(\phi',\kappa')$ that are subjectively more complex than $(\phi,\kappa)$. On the other hand, the green region contains the tasks $(\phi',\kappa')$ that are subjectively simpler than $(\phi,\kappa)$. How we obtained the red region is obvious given our previous theorem. So, let us elaborate on how the green region arises. First of all, $\kappa\geq\kappa'$ and $\phi\geq\phi'$ jointly imply $(\phi,\kappa)\succeq_w(\phi',\kappa')$, again by our theorem. Then, let us consider some $(\phi',\kappa')$ such that $\kappa'<\kappa$ and $\phi'\geq\phi$. By $\phi_w$ being increasing, it follows that $\phi_w(\kappa')<\phi_w(\kappa)<\phi$. Hence, once again by our previous theorem, we obtain $(\phi,\kappa)\succeq_w(\phi',\kappa')$. 

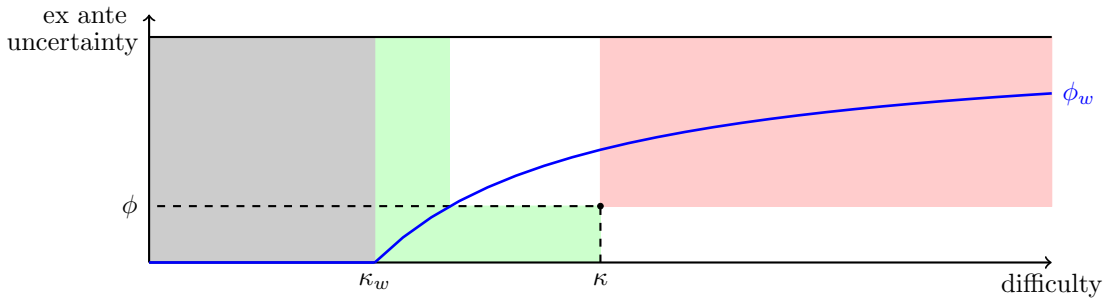
\begin{figure}[h!]
\begin{center}
\begin{tikzpicture}[scale=3]
\filldraw[opaque,white!80!red] (2,0.25) rectangle (4,1);
\filldraw[opaque,white!80!green] (1,0) rectangle (2,0.25);
\filldraw[opaque,white!80!green] (0,0.25) rectangle (1.33,1);
\filldraw[opaque,white!80!black] (0,0) rectangle (1,1);
\draw[line width=0.8,->] (0,0) -- (0,1.1);
\draw (-0.05,1.1) node[left] {\footnotesize{ex ante}};
\draw (0,0.97) node[left] {\footnotesize{uncertainty}};
\draw[line width=0.8,->] (0,0) -- (4,0) node[below] {\footnotesize{difficulty}};
\draw[line width=0.8] (0,1) -- (4,1);
\draw[line width=0.8,dashed] (2,0) node[below] {\footnotesize{$\kappa$}} -- (2,0.25) -- (0,0.25) node[left] {\footnotesize{$\phi$}};
\draw (1,0) node[below] {\footnotesize{$\kappa_w$}};
\filldraw (2,0.25) circle(0.35pt);
\draw[blue,line width=1,domain=0:1] plot(\x,{0});
\draw[blue,line width=1,domain=1:4] plot(\x,{1-1/\x});
\draw[blue] (4,0.75) node[right] {\footnotesize{$\phi_w$}};
\end{tikzpicture}
\end{center}\vspace{-1\baselineskip}
\caption{\footnotesize{This is the case when the ex ante uncertainty is small and therefore it is optimal not to acquire any information before making a guess. Once again, the red area contains the tasks that are deemed more complex than $(\phi,\kappa)$, the green area the tasks that are deemed simpler than $(\phi,\kappa)$, and the grey area the trivial tasks.}}
\label{FIG:main theorem non-binding information condition}
\end{figure}

Let us now focus on the second case (Figure \ref{FIG:main theorem non-binding information condition}), where we take a task $(\phi,\kappa)$ such that $\phi\leq\phi_w(\kappa)$. Once again, by our previous theorem, the red region is the set of tasks that are subjectively more complex than $(\phi,\kappa)$. And once again, the green region is the set of tasks that are subjectively simpler than $(\phi,\kappa)$. Regarding this last part, let us elaborate only on the non-obvious case where $\kappa'<\kappa$ and $\phi'>\phi$. If $\kappa'$ is such that $\phi_w(\kappa')>\phi$, then by our theorem $(\phi,\kappa)$ and $(\phi',\kappa')$ are not comparable via $\succeq_w$. 

\begin{remark}
Note that in most of the existing literature, complexity is taken as a synonym of difficulty. Here we show that this is not the case, as complexity depends both on difficulty and ex ante uncertainty. This is consistent with the idea that a task may be deemed complex because it has not been previously encountered by the agent. For instance, students often complain that they did not have enough practice questions similar to the ones they faced in their final exam, even though their actual exam was not particularly difficult. Nevertheless, even though increased difficulty is not sufficient for increased complexity, it is still a necessary condition, i.e., unless both $(\phi,\kappa)$ and $(\phi',\kappa')$ are both trivial, $(\phi',\kappa')\succeq_w(\phi,\kappa)$ will necessarily imply $\kappa'\geq\kappa$.
\end{remark}

\section{Properties of subjective complexity order}\label{S:properties}

Our definition of subjective complexity is robust with respect to the extrinsic reward. However, such robustness comes at the cost of not being able to rank all tasks. Let us provide some results on the structure of $\succeq_w$ for different $w\geq0$. 

\begin{proposition}\label{P:transitivity incompleteness}
For every $w\geq0$, the relation $\succeq_w$ is transitive and incomplete. 
\end{proposition}

The previous result can be easily illustrated in the two figures above, i.e., there is always a white region which includes the tasks that are subjectively neither more complex nor simpler than any given task. 

Note that within $\mathcal{T}_w$ the relation $\succeq_w$ has the same structure as the incomplete preference relations in \cite{Ok2002}. To understand the connection, recall that in that paper preferences are represented by a vector-valued utility function. In the context of our paper this would mean that there exists some 
$$\boldsymbol{u}:\mathcal{T}_w\rightarrow\mathbb{R}^2$$ 
such that $(\phi,\kappa)\succeq_w (\phi',\kappa')$ if and only if $\boldsymbol{u}_1(\phi,\kappa)\geq \boldsymbol{u}_1(\phi',\kappa')$ and $\boldsymbol{u}_2(\phi,\kappa)\geq \boldsymbol{u}_2(\phi',\kappa')$, where $\boldsymbol{u}_1(\phi,\kappa):=\kappa$ and $\boldsymbol{u}_2(\phi,\kappa):=\min\{\phi_w(\kappa),\phi\}$. The latter also distinguishes our notion of complexity from other definitions in the literature which typically induce a complete order, e.g., the signal-to-noise ratio \citep{Callander2011, FehrRangel2011, Goncalves2024}.

Now, let us look at the comparative statics with respect to the intrinsic incentives. This will provide us some insight on whether we should be using more or less intrinsically motivated agents to compare tasks in terms of complexity.

\begin{proposition}\label{P:more complete}
If $w'>w$, then the following hold:
\begin{itemize}
\item[(a)]  $\mathcal{T}_{w'}\subseteq\mathcal{T}_w$.
\item[(b)] $\succeq_w \ \subseteq \ \succeq_{w'}$.  
\end{itemize}
\end{proposition}

\begin{remark}\label{REM:monotonicity strict relation}
Monotonicity of $\succeq_w$ does not extend to monotonicity of the asymmetric part $\succ_w$ of the relation. Indeed, it might be the case that $(\phi,\kappa)\succ_w(\phi',\kappa')$ for some $w$, and then $(\phi,\kappa)\sim_{w'}(\phi',\kappa')$ for some $w'>w$. This is, for instance, the case when $\kappa=\kappa'$ and $\phi'>\phi_w(\kappa)>\phi>\phi_{w'}(\kappa)$. However, what will never happen is a complete reversal of subjective complexity, i.e., it cannot be that $(\phi,\kappa)\succ_w(\phi',\kappa')$ and $(\phi,\kappa)\prec_{w'}(\phi',\kappa')$.
\end{remark}

Going back to Proposition \ref{P:more complete}, the fact that $\succeq_w$ is increasing in $w$ follows from $\phi_w$ being pointwise decreasing, i.e., if $w'>w$, then $\phi_{w'}(\kappa)\leq\phi_w(\kappa)$ for every $\kappa>0$. The intuition is pretty clear: stronger intrinsic incentives will lead to more frequent information acquisition. At the same time, more tasks will become trivial, and therefore equivalent in terms of complexity. In fact, it is not difficult to see that in the limit, as $w$ becomes arbitrarily large, agent deems all tasks equally complex.

Together, Proposition \ref{P:more complete} and Remark \ref{REM:monotonicity strict relation}, point to an interesting tradeoff. Stronger intrinsic incentives allow us to compare more tasks in terms of complexity, but the differences between tasks become less salient. This is because expected accuracy becomes inflated by the intrinsic incentives, and therefore all expected accuracies are concentrated near extreme beliefs, which makes it more difficult to distinguish the complexity of different tasks.

\section{Subjective complexity and effort}\label{S:complexity and effort}

The relationship between complexity and effort has been studied by several papers in the literature \citep[][and references therein]{Goncalves2024}. In this paper, we naturally assume that expected response time is a proxy for effort, which is in turn positively correlated with the cost $C\bigl(\pi(\cdot|x,w,\phi,\kappa)\bigr)$ that the agent incurs for optimally acquiring information about the task. 

For starters, it is not difficult to see that for any fixed $x\geq x_0$, the induced effort is not always monotonic in complexity. This finding is not surprising, and it has been already pointed out in the literature. Here we will focus on the the problem from a different angle, focusing on robustness with respect to the extrinsic reward, i.e., we ask whether the non-monotonicity can arise for every $x\geq x_0$. 

\begin{theorem}\label{T:monotonicity I}
Fix $w\geq0$, and take an arbitrary $(\phi,\kappa)\in\mathcal{T}_w$ such that $\phi>\phi_w(\kappa)$. Then, there exists $(\phi',\kappa')\succ_w(\phi,\kappa)$ such that $C\bigl(\pi(\cdot|x,w,\phi',\kappa')\bigr)< C\bigl(\pi(\cdot|x,w,\phi,\kappa)\bigr)$ for all $x\geq x_0$.
\end{theorem} 

The previous result shows that for all tasks for which the agent optimally acquires some information, we can always find a more complex task which will induce less effort regardless of the size of the extrinsic reward. In this sense, non-monotonicity of effort with respect to complexity is both generic and robust with respect to the extrinsic incentives. Graphically this is illustrated in Figure \ref{FIG:monotonicity} below. Of course, for extrinsic reward $x_0$ it is pretty obvious, as the agent will not acquire any information when facing $(\phi',\kappa')$, as opposed to when facing $(\phi,\kappa)$. The interesting part arises when the extrinsic reward increases to any $x>x_0$. In this case, although $(\phi',\kappa')$ is harder, it involves lower ex ante uncertainty. And this is exactly what makes it cheaper. 

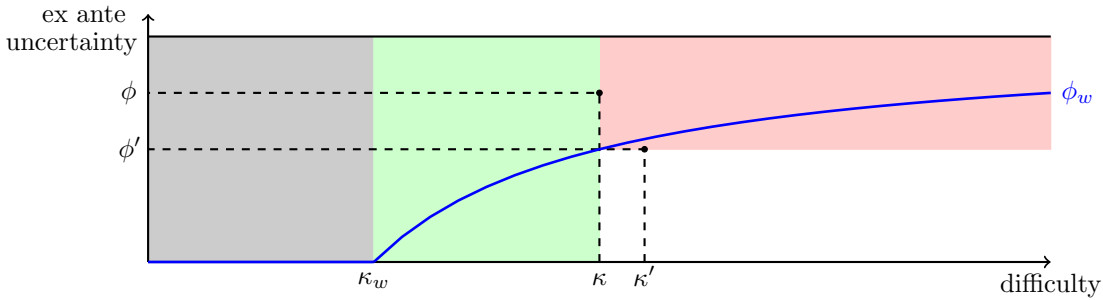
\begin{figure}[h!]
\begin{center}
\begin{tikzpicture}[scale=3]
\filldraw[opaque,white!80!red] (2,0.5) rectangle (4,1);
\filldraw[opaque,white!80!black] (0,0) rectangle (1,1);
\filldraw[opaque,white!80!green] (1,0) rectangle (2,1);
\draw[line width=0.8,->] (0,0) -- (0,1.1);
\draw (-0.05,1.1) node[left] {\footnotesize{ex ante}};
\draw (0,0.97) node[left] {\footnotesize{uncertainty}};
\draw[line width=0.8,->] (0,0) -- (4,0) node[below] {\footnotesize{difficulty}};
\draw[line width=0.8] (0,1) -- (4,1);
\draw[line width=0.8,dashed] (2,0) node[below] {\footnotesize{$\kappa$}} -- (2,0.75) -- (0,0.75) node[left] {\footnotesize{$\phi$}};
\draw[line width=0.8,dashed] (2.2,0) -- (2.2,0.5) -- (0,0.5);
\draw (1,0) node[below] {\footnotesize{$\kappa_w$}};
\draw (2.2,0.03) node[below] {\footnotesize{$\kappa'$}};
\draw (0.02,0.5) node[left] {\footnotesize{$\phi'$}};
\filldraw (2,0.75) circle(0.35pt);
\filldraw (2.2,0.5) circle(0.35pt);
\draw[blue,line width=1,domain=0:1] plot(\x,{0});
\draw[blue,line width=1,domain=1:4] plot(\x,{1-1/\x});
\draw[blue] (4,0.75) node[right] {\footnotesize{$\phi_w$}};
\end{tikzpicture}
\end{center}
\vspace{-1\baselineskip}
\caption{\footnotesize{THere is some $\kappa'>\kappa$ such that task $(\phi',\kappa')$ is strictly more complex than $(\phi,\kappa)$, but at the same time it always induces less effort, regardless of how large the extrinsic reward is.}}
\label{FIG:monotonicity}
\end{figure}

In fact, it is not difficult to show that if complexity is entirely driven by ex ante uncertainty, more complex tasks will always induce more effort. That is, whenever $\phi>\phi'$, it will be the case that $(\phi,\kappa)\succ_w(\phi',\kappa)$, and furthermore $C\bigl(\pi(\cdot|x,w,\phi,\kappa)\bigr)> C\bigl(\pi(\cdot|x,w,\phi',\kappa)\bigr)$ for all $x\geq x_0$. 

Interestingly, while the effect of ex ante uncertainty is clear, the same is not true for difficulty. 

\begin{proposition}\label{P:reversal effort}
For every $\phi$, there exists some $\kappa_\phi>\kappa_w$ such that, for any two $\kappa'>\kappa>\kappa_\phi$:
\begin{eqnarray*}
C\bigl(\pi(\cdot|x,w,\phi,\kappa)\bigr)&>&C\bigl(\pi(\cdot|x,w,\phi,\kappa')\bigr),\\
C\bigl(\pi(\cdot|x',w,\phi,\kappa)\bigr)&<&C\bigl(\pi(\cdot|x',w,\phi,\kappa')\bigr),
\end{eqnarray*}
for extrinsic rewards $x'>x\geq x_0$.
\end{proposition}

The previous result suggests that effort rankings are never robust along the difficulty dimension. Thus, besides the fact that effort is not monotonic with respect to complexity, the strength of the extrinsic reward also matters, in the sense that for relatively small rewards the simpler task induces more effort, whereas for high rewards the more complex task induces more effort. This result can be seen as the robust counterpart of \citet[][Prop. 3]{Goncalves2024}, who shows that for a fixed $x$, effort is not monotonic with respect to difficulty.













	\appendix
	
	\numberwithin{equation}{section}
	
	\setcounter{lemma}{0}
	\setcounter{corollary}{0}
	\setcounter{theorem}{0}
	\setcounter{proposition}{0}
	\setcounter{rema}{0}
	\setcounter{defin}{0}
	
	\renewcommand{\thelemma}{\Alph{section}\arabic{lemma}}
	\renewcommand{\theproposition}{\Alph{section}\arabic{proposition}}
	\renewcommand{\thecorollary}{\Alph{section}\arabic{corollary}}
	\renewcommand{\thetheorem}{\Alph{section}\arabic{theorem}}
	\renewcommand{\therema}{\Alph{section}\arabic{rema}}
	\renewcommand{\thedefin}{\Alph{section}\arabic{defin}}

\section{Proofs}

\begin{proof}[\textbf{\textup{Proof of Proposition \ref{P:trivial task}}}]
Fix some $w\geq0$. By a standard concavification argument, there exists some $\delta(x,w,\kappa)\in\big[0,\nicefrac{1}{2}\big]$ such that the optimal signal distributes all probability between $\delta(x,w,\kappa)$ and $1-\delta(x,w,\kappa)$ whenever $p\in\big(\delta(x,w,\kappa),1-\delta(x,w,\kappa)\big)$, and it is completely uninformative otherwise. Note that $\delta(x,w,\kappa)$ is increasing in $\kappa$ and decreasing in $x$. Then, define 
\begin{equation}\label{EQ:proof of Proposition trivial}
\kappa_w:=\sup\big\{\kappa>0:\delta(x_0,w,\kappa)=0\big\}.
\end{equation}
Note that $\kappa_w=0$ if and only if $x_0=w=0$. 

\vspace{0.5\baselineskip} \noindent $(i)\Rightarrow(ii):$ By construction, for every $\kappa\leq \kappa_w$ and every $x\geq x_0$, it is the case that $\delta(x,w,\kappa)\leq \delta(x_0,w,\kappa)\leq \delta(x_0,w,\kappa_w)=0$. This means that $(\phi,\kappa)$ is trivial given $w$ for every $\phi\in[0,1]$.

\vspace{0.5\baselineskip} \noindent $(ii)\Rightarrow(i):$ Suppose that $\kappa>\kappa_w$. Then, by construction $\delta(x_0,w,\kappa)>0$. Therefore, there is some $\phi>0$ sufficiently close to 0 such that the optimal signal is completely uninformative. This means that $F(x_0,w,\phi,\kappa)<F(x_0,w,\phi,\kappa_w)=1$, i.e., $(\phi,\kappa)$ is not trivial given $w$.

\vspace{0.5\baselineskip} \noindent Finally, by (\ref{EQ:proof of Proposition trivial}) it follows directly that $\kappa_w$ is increasing in $w$.
\end{proof}

\begin{proof}[\textbf{\textup{Proof of Theorem \ref{T:main theorem incompleteness}}}]
Fix the intrinsic incentives $w\geq0$, and take an arbitrary pair $(\phi,\kappa)$. Similarly to the proof of Proposition \ref{P:trivial task}, for every $x\geq x_0$ there exists some $\delta_x\in\big[0,\nicefrac{1}{2}\big]$ such that the optimal signal distributes all probability between $\delta_x$ and $1-\delta_x$ whenever $p\in(\delta_x,1-\delta_x)$, and it is completely uninformative otherwise. Therefore, we obtain 
\begin{equation}\label{EQ:proof main theorem 1}
F(x,w,\phi,\kappa)=\begin{cases}
1-\nicefrac{\phi}{2} & \mbox{if } \phi\leq \phi_w^x,\\
1-\delta_x & \mbox{if } \phi> \phi_w^x,
\end{cases}
\end{equation}
where $\phi_w^x(\kappa):=1-2\big|\delta_x-\nicefrac{1}{2}\big|$ is the threshold of the degree of ex ante uncertainty for acquiring information. Then, define $\phi_w(\kappa):=\phi_w^{x_0}(\kappa)$. 

\vspace{0.5\baselineskip} \noindent $(ii)\Rightarrow(i):$ Take $\kappa'\geq\kappa$ and $\phi'\geq \min\{\phi_w(\kappa),\phi\}$, and consider two cases:

\vspace{0.5\baselineskip} \noindent First, let $\phi'\geq \phi$. Note that Equation (\ref{EQ:proof main theorem 1}) can be equivalently rewritten as
\begin{equation}\label{EQ:proof main theorem 4}
F(x,w,\phi,\kappa)=\max\{1-\nicefrac{\phi}{2},1-\delta_x\}.
\end{equation}
Letting $\delta_x'$ be the low posterior we obtain from the concavification exercise given the parameters $(\phi',\kappa')$, it is not difficult to verify that $\delta_x\leq \delta_x'$. Hence, it follows directly from (\ref{EQ:proof main theorem 4}) that $F(x,w,\phi',\kappa')\leq F(x,w,\phi,\kappa)$ for all $x\geq x_0$, and a fortiori we obtain $(\phi',\kappa')\succeq_w(\phi,\kappa)$.

\vspace{0.5\baselineskip} \noindent Second, let $\phi>\phi'\geq \phi_w(\kappa)$. By definition, $\phi_w^x(\kappa)$ is decreasing in $x$, and hence $\phi>\phi_w^x(\kappa)$. So, combined with (\ref{EQ:proof main theorem 1}), we obtain $F(x,w,\phi,\kappa)=F(x,w,\phi',\kappa)=1-\delta_x$. Moreover, as shown in the first case above, $F$ is decreasing in $\kappa$ and therefore $F(x,w,\phi',\kappa')\leq F(x,w,\phi',\kappa)$. Combining the two yields $F(x,w,\phi',\kappa')\leq F(x,w,\phi,\kappa)$ for all $x\geq x_0$, and a fortiori we obtain $(\phi',\kappa')\succeq_w(\phi,\kappa)$.

\vspace{0.5\baselineskip} \noindent $(i)\Rightarrow(ii):$ Suppose that $\kappa_w<\kappa'<\kappa$. Then, there exists some $x\geq x_0$ such that $\phi>\phi_w^x(\kappa)$, and therefore $F(x,w,\phi,\kappa)=1-\delta_x$. But then, by (\ref{EQ:proof main theorem 4}), it is also the case that $F(x,w,\phi',\kappa')\geq1-\delta_x'$. And by $\kappa'<\kappa$, we have $1-\delta_x'>1-\delta_x$, and a fortiori $F(x,w,\phi',\kappa')>F(x,w,\phi,\kappa)$, meaning that $(\phi',\kappa')\nsucceq_w(\phi,\kappa)$.

\vspace{0.5\baselineskip} \noindent Finally, suppose that $\kappa_w<\kappa\leq\kappa'$ and $\phi'<\min\{\phi,\phi_w(\kappa)\}$. The latter implies that $\phi'<\phi$. Since both $(\phi,\kappa)$ and $(\phi',\kappa')$ are non-trivial given $w$, it means that $F(x,w,\phi,\kappa)=1-\nicefrac{\phi}{2}<1-\nicefrac{\phi'}{2}=F(x,w,\phi,\kappa)$, and therefore $(\phi',\kappa')\nsucceq_w(\phi,\kappa)$.
\end{proof}

\begin{proof}[\textup{\textbf{Proof of Proposition \ref{P:transitivity incompleteness}}}]
\textsc{Transitivity:} It follows directly from Theorem \ref{T:main theorem incompleteness}. Namely, begin with $(\phi'',\kappa'')\succeq_w(\phi',\kappa')\succeq_w(\phi,\kappa)$. Hence, we have $\kappa''\geq\kappa'\geq\kappa$ and $\phi''\geq\min\{\phi',\phi_w(\kappa')\}$ and $\phi'\geq\min\{\phi,\phi_w(\kappa)\}$. Then, we consider two cases: 

\vspace{0.2\baselineskip} \noindent (i) $\phi''\geq\phi':$ It follows directly $\phi''\geq\min\{\phi,\phi_w(\kappa)\}$, and therefore $(\phi'',\kappa'')\succeq_w(\phi,\kappa)$.

\vspace{0.2\baselineskip} \noindent (i) $\phi'>\phi''\geq\phi_w(\kappa'):$ By monotonicity of $\phi_w$ we have $\phi_w(\kappa')\geq\phi_w(\kappa)\geq\min\{\phi,\phi_w(\kappa)\}$. Hence, we obtain $\phi''\geq\min\{\phi,\phi_w(\kappa)\}$, and a fortiori $(\phi'',\kappa'')\succeq_w(\phi,\kappa)$.

\vspace{0.2\baselineskip} \noindent \textsc{Incompleteness:} By the fact that $\phi_w$ approaches 1 in the limit, there exists some $\kappa>\kappa_w$ such that $\phi_w(\kappa)>0$. Then, consider $\phi>\phi_w(\kappa)>\phi'$, and $\kappa'>\kappa$. By Theorem \ref{T:main theorem incompleteness}, we obtain $(\phi,\kappa)\nsucceq_w(\phi',\kappa')$ and $(\phi',\kappa')\nsucceq_w(\phi,\kappa)$, meaning that $\succeq_w$ is incomplete.
\end{proof}

\begin{proof}[\textup{\textbf{Proof of Proposition \ref{P:more complete}}}]
(a) By construction, if $w'>w$ then $\phi_{w'}(\kappa)\leq\phi_w(\kappa)$ for every $\kappa>0$. This means first of all that $\mathcal{T}_{w'}\subseteq \mathcal{T}_w$. 

\vspace{0.5\baselineskip} \noindent (b) Let $w'>w$, and take some $(\phi',\kappa')\notin\mathcal{T}_{w'}$ such that $(\phi,\kappa)\succeq_w(\phi',\kappa')$. This means that $\kappa\geq\kappa'$ and $\phi\geq\min\{\phi',\phi_w(\kappa')\}$. Hence, we also have $\phi\geq\min\{\phi',\phi_{w'}(\kappa')\}$, and therefore $(\phi,\kappa)\succeq_{w'}(\phi',\kappa')$.
\end{proof}

\begin{proof}[\textbf{\textup{Proof of Theorem \ref{T:monotonicity I}}}]
Let $\phi':=\phi_w(\kappa)$. Denote by $p$ and $p'$ two arbitrary priors that correspond to $\phi$ and $\phi'$ respectively. Then, for every $x\geq x_0$, we obtain 
\begin{equation}\label{EQ:proof of theorem monotonicity 1}
C\bigl(\pi(\cdot|x,w,\phi,\kappa)\bigr)- C\bigl(\pi(\cdot|x,w,\phi',\kappa)\bigr)=\kappa\big(c(p)-c(p')\bigr)=\varepsilon>0.
\end{equation}
Take some $\kappa'>\kappa$ such that $(\kappa'-\kappa)\big(c(0)-c(\nicefrac{1}{2})\bigr)=\varepsilon$. Then, for any $x\geq x_0$, we obtain
\begin{equation}\label{EQ:proof of theorem monotonicity 2}
C\bigl(\pi(\cdot|x,w,\phi',\kappa')\bigr)- C\bigl(\pi(\cdot|x,w,\phi',\kappa)\bigr)=\kappa'\big(c(\delta_x')-c(p')\bigr)-\kappa\big(c(\delta_x)-c(p')\bigr)<\varepsilon,
\end{equation}
where $\delta_x$ and $\delta_x'$ are the posteriors that we obtain from the concavification exercise, like in the proof of Theorem \ref{T:main theorem incompleteness}. The last inequality follows from $c(\delta_x),c(\delta_x')\leq c(0)$ and $c(p')>c(\nicefrac{1}{2})$. Hence, by (\ref{EQ:proof of theorem monotonicity 1}) and (\ref{EQ:proof of theorem monotonicity 2}), we obtain $C\bigl(\pi(\cdot|x,w,\phi',\kappa')\bigr)- C\bigl(\pi(\cdot|x,w,\phi,\kappa)\bigr)<0$.
\end{proof}

\begin{proof}[\textbf{\textup{Proof of Proposition \ref{P:reversal effort}}}]
Let $\phi<1$, and define $\kappa_\phi:=\phi_w^{-1}(\phi)$. Fix $\kappa'>\kappa>\kappa_\phi$, and observe that there is some $x>x_0$ such that $\pi(\cdot|x,w,\phi,\kappa')$ is completely uninformative as opposed to $\pi(\cdot|x,w,\phi,\kappa)$ which provides some information. Therefore, we obtain $C\bigl(\pi(\cdot|x,w,\phi,\kappa)\bigr)>C\bigl(\pi(\cdot|x,w,\phi,\kappa')\bigr)$. Moreover, for each $x''\geq0$, denote by $\delta_x$ and $\delta_x'$ the low posteriors in the support of $\pi(\cdot|x'',w,\phi,\kappa)$ and $\pi(\cdot|x'',w,\phi,\kappa')$ respectively, like in the proof of Theorem \ref{T:main theorem incompleteness}. Note that, as $x''$ becomes arbitrarily large, both $\delta_x$ and $\delta_x'$ converge to 0. Thus, by continuity of $c$, there exists some $x'$ such that  $C\bigl(\pi(\cdot|x',w,\phi,\kappa)\bigr)<C\bigl(\pi(\cdot|x',w,\phi,\kappa')\bigr)$.
\end{proof}

\bibliographystyle{apalike}

\bibliography{literature}

\begin{thebibliography}{}

\bibitem[Abreu and Rubinstein, 1988]{AbreuRubinstein1988}
Abreu, D. and Rubinstein, A. (1988).
\newblock The structure of nash equilibria in repeated games with finite
  automata.
\newblock {\em Econometrica}, 56:1259--1281.

\bibitem[Alaoui and Penta, 2022]{AlaouiPenta2022}
Alaoui, L. and Penta, A. (2022).
\newblock Cost-benefot analysis in reasoning.
\newblock {\em Journal of Political Economy}, 130(4).

\bibitem[Anscombe and Aumann, 1963]{AnscombeAumann1963}
Anscombe, F. and Aumann, R. (1963).
\newblock A definition of subjective probability.
\newblock {\em Annals of Mathematical Statistics}, 34:199--205.

\bibitem[Banovetz and Oprea, 2023]{BanovetzOprea2023}
Banovetz, J. and Oprea, R. (2023).
\newblock Complexity and procedural choice.
\newblock {\em American Economic Journal: Microeconomics}, 15:3913--3951.

\bibitem[Brier, 1950]{Brier1950}
Brier, G. (1950).
\newblock Verification of forecasts expressed in terms of probability.
\newblock {\em Monthly Weather Review}, 78:1--3.

\bibitem[Callander, 2011]{Callander2011}
Callander, S. (2011).
\newblock Searching and learning by trial and error.
\newblock {\em American Economic Review}, 111:2277--2308.

\bibitem[Caplin, 2025]{Caplin2025}
Caplin, A. (2025).
\newblock {\em An introduction to cognitive economics: The science of
  mistakes}.
\newblock Palgrave.

\bibitem[Caplin et~al., 2022]{Caplin2022}
Caplin, A., Dean, M., and Leahy, J. (2022).
\newblock Rationally inattentive behavior: Characterizing and generalizing
  {Shannon} entropy.
\newblock {\em Journal of Political Economy}, 130:1676--1715.

\bibitem[Cover and Thomas, 2006]{CoverThomas2006}
Cover, T. and Thomas, J. (2006).
\newblock {\em Elements of Information Theory}.
\newblock Wiley-Interscience.

\bibitem[Dean and Neligh, 2023]{DeanNeligh2023}
Dean, M. and Neligh, N. (2023).
\newblock Experimental tests of rational inattention.
\newblock {\em Journal of Political Economy}, 131:3415--3461.

\bibitem[Denti, 2022]{Denti2022}
Denti, T. (2022).
\newblock Posterior separable cost of information.
\newblock {\em American Economic Review}, 112:3215--3259.

\bibitem[Enke, 2024]{Enke2024}
Enke, B. (2024).
\newblock The cognitive turn in behavioral economics.
\newblock {\em Working Paper}.

\bibitem[Enke and Graeber, 2023]{EnkeGraeber2023}
Enke, B. and Graeber, T. (2023).
\newblock Cognitive uncertainty.
\newblock {\em Quarterly Journal of Economics}, 138:2021--2067.

\bibitem[Enke et~al., 2024a]{EnkeGraeberOprea2024}
Enke, B., Graeber, T., and Oprea, R. (2024a).
\newblock Complexity and time.
\newblock {\em Journal of the European Economic Association}, forthcoming.

\bibitem[Enke et~al., 2024b]{EnkeGraeberOpreaYang2024}
Enke, B., Graeber, T., Oprea, R., and Yang, J. (2024b).
\newblock Behavioral attenuation.
\newblock {\em Working Paper}.

\bibitem[Fehr and Rangel, 2011]{FehrRangel2011}
Fehr, E. and Rangel, A. (2011).
\newblock Neuroeconomic foundations of economic choice -- recent advances.
\newblock {\em Journal of Economic Perspectives}, 25:3--30.

\bibitem[Gabaix and Graeber, 2024]{GabaixGraeber2024}
Gabaix, X. and Graeber, T. (2024).
\newblock The complexity of economic decisions.
\newblock {\em Working Paper}.

\bibitem[Goncalves, 2024]{Goncalves2024}
Goncalves, D. (2024).
\newblock Speed, accuracy, and complexity.
\newblock {\em Working Paper}.

\bibitem[H\'{e}bert and Woodford, 2021]{HebertWoodford2021}
H\'{e}bert, B. and Woodford, M. (2021).
\newblock Neighborhood-based information costs.
\newblock {\em American Economic Review}, 111:3225--3255.

\bibitem[Kamenica and Gentzkow, 2011]{KamenicaGentzkow2011}
Kamenica, E. and Gentzkow, M. (2011).
\newblock Bayesian persuasion.
\newblock {\em American Economic Review}, 101:2590--2615.

\bibitem[Mat{\v{e}}jka and McKay, 2015]{Matejka2015}
Mat{\v{e}}jka, F. and McKay, A. (2015).
\newblock Rational inattention to discrete choices: A new foundation for the
  multinomial logit model.
\newblock {\em American Economic Review}, 105:272--298.

\bibitem[Ok, 2002]{Ok2002}
Ok, E. (2002).
\newblock Utility representation of an incomplete preference relation.
\newblock {\em Journal of Economic Theory}, 104:429--449.

\bibitem[Oprea, 2020]{Oprea2020}
Oprea, R. (2020).
\newblock What makes a rule complex?
\newblock {\em American Economic Review}, 110:3913--3951.

\bibitem[Oprea, 2024a]{OpreaReview2024}
Oprea, R. (2024a).
\newblock Complexity and its measurement.
\newblock {\em Handbook of Experimental Methods in the Social Sciences},
  forthcoming.

\bibitem[Oprea, 2024b]{Oprea2024}
Oprea, R. (2024b).
\newblock Decisions under risk are decisions under complexity.
\newblock {\em American Economic Review}, 114:3789--3811.

\bibitem[Rubinstein, 1986]{Rubinstein1986}
Rubinstein, A. (1986).
\newblock Finite automata play the repeated prisoner's dilemma.
\newblock {\em Journal of Economic Theory}, 39:83--96.

\bibitem[Savage, 1954]{Savage1954}
Savage, L. (1954).
\newblock {\em The foundations of statistics}.
\newblock Wiley.

\bibitem[Savage, 1971]{Savage1971}
Savage, L. (1971).
\newblock Elicitation of personal probabilities and expectations.
\newblock {\em Journal of the American Statistical Association}, 66:783--801.

\bibitem[Shorrocks, 1980]{Shorrocks1980}
Shorrocks, A. (1980).
\newblock The class of additively decomposable inequality measures.
\newblock {\em Econometrica}, 48:613--625.

\bibitem[Sims, 2003]{Sims2003}
Sims, C. (2003).
\newblock Implications of rational inattention.
\newblock {\em Journal of Monetary Economics}, 50:665--690.

\bibitem[van~der Wel and van Steenbergen, 2018]{vanderWelvanSteenbergen2018}
van~der Wel, P. and van Steenbergen, H. (2018).
\newblock Pupil dilation as an index of effort in cognitive control tasks: A
  review.
\newblock {\em Psychonomic Bulletin and Review}, 25:2005--2015.

\bibitem[Woodford, 2020]{Woodford2020}
Woodford, M. (2020).
\newblock Modeling imprecision perception, valuation and choice.
\newblock {\em Annual Review of Economics}, 12.

\end{thebibliography}

\end{document}